\documentclass[aps,pre,preprint,superscriptaddress]{revtex4-1}
\usepackage{graphicx,amsmath,amsfonts,amssymb}
\begin{document}

%\preprint{}

\title{Multi-rhythmicity in an optoelectronic oscillator with large delay}
\author{Lionel Weicker}
\email{lweicker@ulb.ac.be}
\affiliation{Optique Nonlin\'{e}aire Th\'{e}orique, Universit\'{e} Libre de Bruxelles, Campus Plaine, CP\ 231, 1050 Bruxelles, Belgium}
\affiliation{Applied Physics Research Group (APHY), Vrije Universiteit Brussel, 1050 Brussel, Belgium}
\affiliation{OPTEL Research Group, CentraleSupélec, LMOPS (EA 4423), 2 rue ´
Edouard Belin, 57070 Metz, France}

\author{Thomas Erneux}
\affiliation{Optique Nonlin\'{e}aire Th\'{e}orique, Universit\'{e} Libre de Bruxelles, Campus Plaine, CP\ 231, 1050 Bruxelles, Belgium}

\author{David P. Rosin}
\affiliation{Department of Physics, Duke University, Durham, North Carolina 27708, USA}

\author{Daniel J. Gauthier}
\affiliation{Department of Physics, Duke University, Durham, North Carolina 27708, USA}


\date{\today}

\begin{abstract}
An optoelectronic oscillator exhibiting a large delay in its feedback loop is
studied both experimentally and theoretically. We show that multiple
square-wave oscillations may coexist for the same values of the parameters
(multi-rhythmicity). Depending on the sign of the phase shift, these regimes
admit either periods close to an integer fraction of the delay or periods
close to an odd integer fraction of twice the delay. These periodic solutions
emerge from successive Hopf bifurcation points and stabilize at a finite
amplitude following a scenario similar to Eckhaus instability in spatially
extended systems.\ We find quantitative agreements between experiments and
numerical simulations.\ The linear stability of the square-waves is
substantiated analytically by determining stable fixed points of a map.
\end{abstract}

%\keywords{}

\maketitle


% If in two-column mode, this environment will change to single-column
% format so that long equations can be displayed. Use
%\begin{widetext}
%\end{widetext}

\section{Introduction\label{intro}}

Nonlinear delay dynamics have been a particularly prolific area of research
in the field of photonic devices during the last 30 years
\cite{erneux2010laser}. A large variety of setups exhibiting optical or
electro-optical delayed feedback loops have been explored for novel
applications, but also as experimental tools for delay systems in general.
They have stimulated fruitful interactions with researchers working in
different fields by emphasizing specific delay-induced phenomena
\cite{erneux,bala,Stepan28032009,atay2010complex,just2010delayed,KalmarNagy01062010,lakshmanan2011dynamics,smith2011introduction}%
.\ Examples include different forms of oscillatory instabilities,
stabilization techniques using a delayed feedback, and synchronization
mechanisms for delay-coupled systems. Most of the current lasers used in
applications are semiconductor lasers (SLs), which are highly sensitive to
optical feedback \cite{soriano2013complex}. Here, the light coming from the
laser is reflected back to the laser after a substantial delay.\ Another popular delay system is an optoelectronic oscillator (OEO) \cite{Larger28092013,devgan2013review} that consists of a laser injecting its light into an optoelectronic loop. For
OEOs, the feedback exhibits a large delay because of a long optical fiber line in the
OEO closed-loop configuration. An OEO is capable of generating, within the same
optoelectronic cavity, either an ultra-low-jitter single-tone microwave
oscillation, as used in radar applications \cite{yanne:09}, or a broadband
chaotic carrier typically intended for physical data encryption in high bit
rate optical communications \cite{gastaud04,callan}. The OEO is a particularly attractive system because it allows quantitative comparisons between experiments and theory
\cite{PhysRevE.79.026208,Levy:09,weicker:12b}.

For systems exhibiting a Hopf bifurcation in the absence of delay, a feedback
with a large delay may lead to the coexistence of stable periodic solutions in
the vicinity of the first Hopf bifurcation point.\ This multi-rhythmicity was
predicted theoretically using a Hopf normal-form equation with a delayed
feedback \cite{PhysRevLett.96.220201}, where the bifurcation scenario is
similar to Eckhaus instability in spatially extended systems
\cite{Tuckerman199057}. The bandpass OEO without its optical fiber line admits
a Hopf bifurcation.\ In this paper, we investigate the stabilization of nearby Hopf
bifurcation branches in this regime.

We conduct here a systematic experimental and numerical study of an OEO
exhibiting a large delay. We show that an OEO
admits coexisting stable periodic square-waves. Depending on the feedback
phase, they are characterized by frequencies close to either $(1+2n)/(2\tau
_{D})$ $(n=0,1,2,...)$ or $n/\tau_{D}$ $(n=1,2,...)$ where $\tau_{D}$ is the
delay of the feedback loop. In order to induce these periodic solutions, we
inject a periodic electrical signal into the oscillator during the
initialization phase of the experiment and then observe the resulting dynamics
after the injected signal is removed. In the simulations, we choose different
initial periodic functions in order to determine different periodic solutions.

Periodic regimes of an OEO showing frequencies that are multiple of
$1/\tau_{D}$ were found in the past. In \cite{Illing2005180,illing06}, the
authors progressively increased the delay and investigated the sequential jump
to stable oscillations of frequency $(2n+1)/(2\tau_{D})$ $(n=0,1,...)$. In
\cite{Rosin11,Rosin_master_thesis}, the authors found numerically periodic
solutions of frequency close to $n/\tau_{D}.$ In this paper, we demonstrate
the multi-rhythmicity phenomenon by exciting square-waves with a specific
frequency (specific $n$). Furthermore, we relate these periodic solutions to
nearby Hopf bifurcation points, a prerequisite for an Eckhaus bifurcation scenario.

\begin{figure}
\centering
\includegraphics{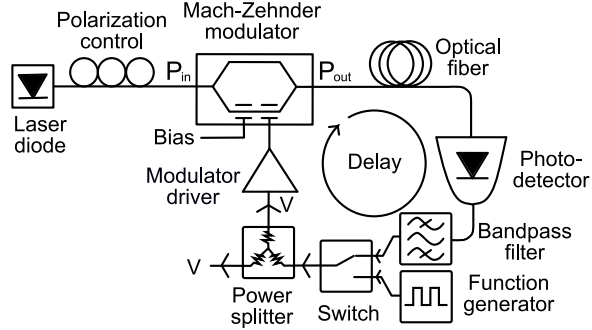}
\caption{Schematic of the experimental setup of an optoelectronic oscillator}
\label{setup_oeo}
\end{figure}

The experimental setup of an OEO is sketched in Fig.\ \ref{setup_oeo}. A semiconductor laser beam is injected into a Mach-Zehnder intensity modulator
(MZM). The MZM induces a nonlinear function of the applied voltage. The
modulated light passes through an optical fiber, which is used as a delay
line, and is injected into an inverting photodetector, which converts the
signal into the electrical domain.\ The voltage emitted from the
photodectector passes through a bandpass filter and then through a power
splitter. Half of the voltage, denoted by $V$, is amplified by an inverting
modulator driver (MD). This electric signal is then reinjected inside the MZM
via its radio frequency input port to close the feedback loop. The voltage
coming out of the other port of the power splitter is used to measure the
dynamical variable $V$ with a high-speed oscilloscope. The device used has an
8 GHz analog bandwidth and a 40 GS/s sampling rate. In our experiments, the
delay of the feedback loop is fixed at $\tau_{D}=22$ ns. The system described
is the same as in Refs.\ \cite{illing06,Rosin11,PhysRevLett.95.203903} except
that a pattern generator has been included to perturb the dynamics of the
system. An electrical switch is used to isolate this pattern generator from the rest of the system. The switch also allows a controllable electrical signal to be combined
with $V$ at the input of the MD.

Mathematically, we consider the evolution equations formulated in Refs.
\cite{callan,Rosin11} with time measured in units of the delay.\ They are
given by \footnote{From Eqs. (1) and (2) in \cite{callan}, we obtain our Eqs. (1) and (2) with $\varepsilon
 = (T\Delta)^{-1} = (T(\omega_{+} - \omega_{-}))^{-1}$, and $\delta=T\omega_{0}^{2}/\Delta=T(\omega_{+}\omega_{-})/(\omega_{+}-\omega_{-})$ and $\beta=\gamma$. $T$ is the delay of the feedback loop, $\omega_{-}$ and $\omega_{+}$ represent the low and high frequency cut-off of the bandpass filter, respectively. Since $\omega_{+} \gg \omega_{-}$, $\varepsilon \simeq (T \omega_{+})^{-1}$ and $\delta \simeq T\omega_{-}$}%
\begin{align}
\varepsilon\frac{dx}{ds} & = -x-\delta y+\beta\left[  \cos^{2}\left(
m+\tanh\left(  x\left(  s-1\right)  \right)  \right)  -\cos^{2}\left(
m\right)  \right]  ,\label{Eq6}\\
\frac{dy}{ds} & = x,\label{Eq7}%
\end{align}
where $s\equiv t/\tau_{D}$ and $x$ is the normalized voltage of the electrical
signal in the OEO. The feedback amplitude $\beta$ and the phase shift $m$ are
two control parameters. $\varepsilon\simeq(\tau_{D}\omega_{+})^{-1}=0.0157$ and $\delta\simeq\tau_{D}\omega_{-}=0.2042$ [26] are
dimensionless time constants fixed by the low and high cut-off frequencies of the bandpass filter denoted by $\omega_{-}$ and $\omega_{+}$, respectively.\ Equations (\ref{Eq6}) and (\ref{Eq7}) are the same equations studied
in \cite{PhysRevLett.95.203903} except of the hyperbolic tangent function in
Eq. (\ref{Eq6}) that accounts for the amplifier saturation.

Equations (\ref{Eq6}) and (\ref{Eq7}) admit a single steady state
$(x,y)=(0,0)$ and its linear stability has been analyzed in detail in Refs.
\cite{Illing2005180,PhysRevLett.95.203903}. Of particular interest are the
primary Hopf bifurcation points, which can be classified into two different
families. In the limit $\delta\rightarrow0$ and $\varepsilon\rightarrow0,$ the
critical feedback amplitudes and the Hopf bifurcation frequencies approach the
limits \footnote{From Eqs. (4) and (5) in \cite{PhysRevLett.95.203903},
introduce $\omega\longrightarrow\omega R,$ $\delta\longrightarrow\varepsilon
R,$ $\varepsilon\longrightarrow1/R.$ The new parameters $\varepsilon$ and
$\delta$ are small and their values are given in Section \ref{intro}.}%
\begin{align}
m >0 :\text{ }\beta_{n} & = 1/\sin(2m),\text{ and }\omega_{n}=(1+2n)\pi\text{
\ }(n=0,1,2,..),\label{OEO9} \\
m <0 :\text{ }\beta_{n} & = -1/\sin(2m),\text{ }\omega_{0}=\sqrt{\delta},\text{
and }\omega_{n}=2n\pi\text{ \ }(n=1,2,..).\label{OE010}%
\end{align}
If $m>0$, the frequencies are odd multiples of $\pi$, meaning that the
successive Hopf bifurcations lead to $2/(1+2n)-$periodic solutions [$2\tau
_{D}/(1+2n)-$periodic solutions in physical time]. If $m<0$ and $n=1,2,..,$
the frequencies are even multiples of $\pi$ and the successive Hopf
bifurcations lead to $1/n-$periodic solutions ($\tau_{D}/n-$periodic solutions
in physical time). In addition, there exists for $m<0$ a Hopf bifurcation
characterized by the low frequency $\omega_{0}=\sqrt{\delta}\ll1$. It leads to
oscillations with a large period compared to $1$ (large period compared to
$\tau_{D}$ in physical time).

The organization of the paper is as follows. In Section \ref{observations}, we
describe the experimental observations and numerical simulations for the two
families of Hopf bifurcations. In Section \ref{stability}, we propose a
partial stability analysis of the plateaus by associating their mean values to
stable fixed points of a map.\ Finally, we discuss our main results in Section
\ref{conclusions}.

\section{Experiments and simulations\label{observations}}

From the linear stability analysis of the zero solution discussed above, we
find that there exist two families of Hopf bifurcations depending on the sign
of $m.$\ For each case, we describe our experimental observations and compare
them to numerical simulations of Eqs. (\ref{Eq6}) and (\ref{Eq7}).

\subsection{Case $m>0$}

\begin{figure}
\centering
\includegraphics[width=10cm]{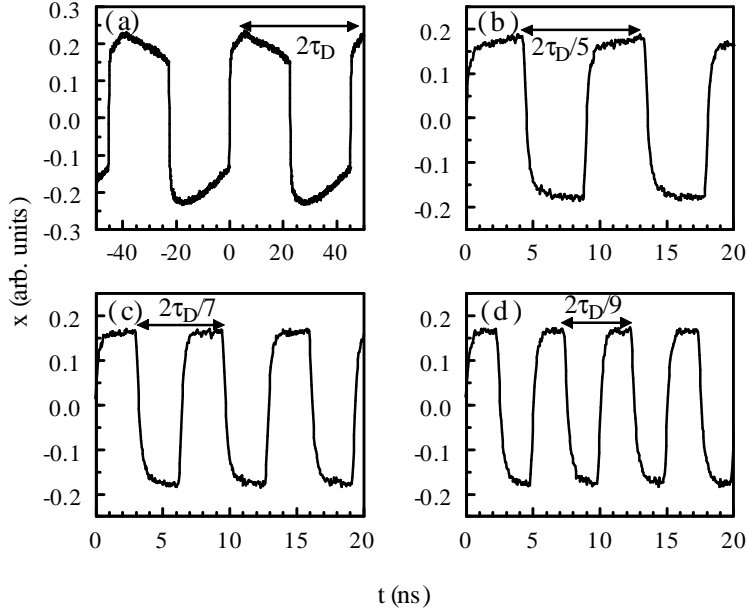}
\caption{Experimental time series obtained after injecting different periodic
signals into the OEO loop during a few seconds and after the injected signal
is removed. The measured values of the parameters are $m=0.665$, $\beta=1.94$,
and $\tau_{D}=22$ ns.}
\label{Fig2}
\end{figure}

If $m>0$, oscillations of period close to $2\tau_{D}$ (corresponding to a
frequency of $22.7$ MHz) are observed experimentally [see Fig.
\ref{Fig2}(a)]. In order to find harmonic oscillations, we excite the system
with signals at different frequencies. To this end, the pattern generator
injects different periodic signals into the OEO loop during a few seconds.
Figure\ \ref{Fig2}(b) shows square-wave oscillations of period close to
$2\tau_{D}/5$ obtained by injecting a square-wave signal of frequency $114$
MHz. Similarly, by exciting the OEO with sine-wave signals of frequency $159$
MHz and of frequency $205$ MHz, we obtain $2\tau_{D}/7$ and $2\tau_{D}%
/9-$periodic oscillations, respectively [Fig. \ref{Fig2}(c) and Fig.
\ref{Fig2}(d), respectively]. $2\tau_{D}/3-$periodic oscillations are also
observed but are not shown for clarity.\ The observation of stable
oscillations characterized by higher frequencies ($n>4$) is not possible
because of the bandwidth limitation of the pattern generator.%

\begin{figure}
\centering
\includegraphics[width=10cm]{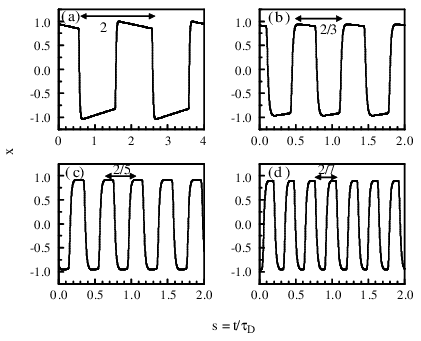}
\caption{Numerical time series obtained from Eqs. (\ref{Eq6}) and (\ref{Eq7}).
(a) Oscillations of period close to $2$ with $x\left(  s\right)  =\cos\left(
\pi s\right)  $ and $y\left(  s\right)  =0$ $(-1<s<0);$ (b) oscillations of
period close to $2/3$ with $x\left(  s\right)  =\cos\left(  3\pi s\right)  $
and $y\left(  s\right)  =0$ $(-1<s<0);$ (c) oscillations of period close to
$2/5$ with $x\left(  s\right)  =\cos\left(  5\pi s\right)  $ and $y\left(
s\right)  =0$ $(-1<s<0)$; (d) oscillations of period close to $2/7$ with
$x\left(  s\right)  =\cos\left(  7\pi s\right)  $ and $y\left(  s\right)  =0$
$(-1<s<0)$. The values of the control parameters are the same as in Fig.
\ref{Fig2}: $m=0.665$ and $\beta=1.94$. }
\label{Fig3}
\end{figure}

We next integrate numerically Eqs. (\ref{Eq6}) and (\ref{Eq7}) using the same
values of the parameters as for the experiments. Figure \ref{Fig3} shows four
different time series obtained using different initial functions described in
the caption. We note that the shape and the period of the oscillations are in
good agreement with the experimental observations. The plateaus of the
square-wave are slightly increasing or decreasing in time, which is an effect
of the small parameters $\varepsilon$ and $\delta$.\ If we decrease their
values, the plateaus become flatter. Another point raised by the numerical
simulations and by the experimental observations is that the mean values of
the plateaus are roughly the same for the main and harmonic periodic
solutions. From Fig. \ref{Fig3}(d), we evaluate theses values as%
\begin{equation}
x_{\max}\simeq0.9\text{ and }x_{\min}\simeq-0.95.\label{extrema}%
\end{equation}

$2/9-$periodic oscillations are also found numerically but they are unstable
for long time. They are stable if we slightly decrease $\varepsilon$. A smaller $\varepsilon$ leads to sharper transition layers and contribute to the overal stability of the square-wave. The discrepancy between experimental and numerical solutions for the $2/9$-periodic regimes could be the result that the model slightly overestimated the effect of the amplifier saturation (value of $d$) which contributes to smooth the transition layers.

\begin{figure}
\centering
\includegraphics[width=10cm]{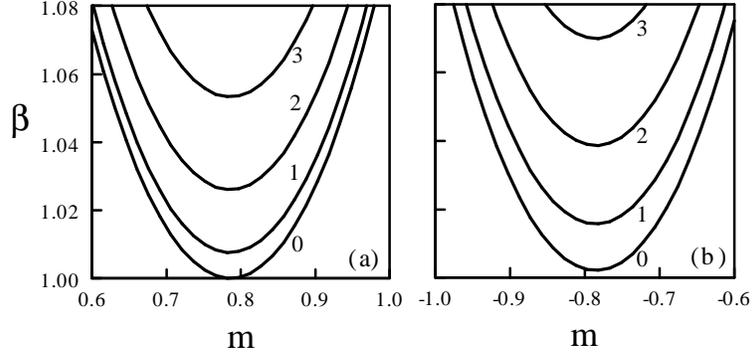}
\caption{Hopf bifurcation lines in the $(m,\beta)$ parameter plane. (a) corresponds to the case $m$ positive and (b) to the case $m$ negative. They have been determined numerically from the exact conditions with $\varepsilon=0.0157$ and $\delta=0.2042$. The numbers in the figures indicate the value of $n$ corresponding to a specific frequency defined in (\ref{OEO9}) and (\ref{OE010}). All curves are nearly parabolic with a minimum at $m=\pm\pi/4$. If $\varepsilon\rightarrow0$ and $\delta\rightarrow0$, all curves moves to a unique parabola with minimum located at $(m,\beta)=(\pm\pi/4,1)$.}
\label{Duke100}
\end{figure}

We also examine the effect of changing $m$. Figure \ref{Duke100}(a) shows the first Hopf bifurcation lines in the $(m,\beta)$ parameter plane. There is a stable steady state if
$\beta<1$.\ Increasing $\beta$ leads to a critical point $\beta_{H1}>1$ where
oscillations of period $2$ appear. The minimal value of $\beta_{H1}$ is
obtained if $m=\pi/4$. By progressively increasing $\beta$ from $\beta_{H1}$,
we may generate stable higher-order harmonic oscillations that become more
robust with respect to small perturbations.

\subsection{Case $m<0$}

\begin{figure}
\centering
\includegraphics[width=10cm]{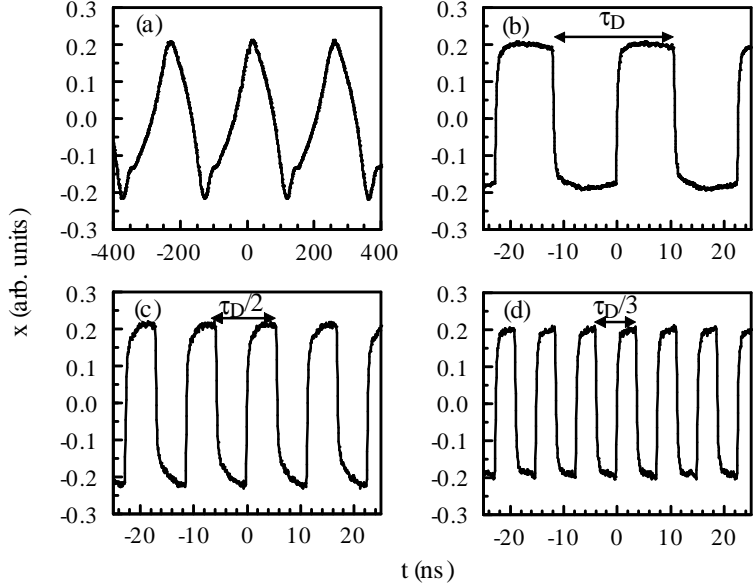}
\caption{Experimental time series obtained after injecting different periodic
signals into the OEO loop during a few seconds and after the injected signal
is removed. (a) Low-frequency oscillations of period close to $12\tau_{D}$
obtained by injecting a sine-wave signal of frequency $5$ MHz; (b)
oscillations of period close to $\tau_{D}$ obtained by injecting a sine-wave
signal of frequency $45.5$\ MHz; (c) oscillations of period close to $\tau
_{D}/2$ obtained by injecting a sine-wave signal of frequency $90.9$ MHz; (d)
oscillations of period close to $\tau_{D}/3$ obtained by injecting a sine-wave
signal of frequency $136$ MHz. The values of the delay is $\tau_{D}=22$
ns.\ The measured values of the control parameters are $m=-0.845$ and
$\beta=1.94$ for (a) and (b) and $m=-0.785$ and $\beta=2.2$ for (c) and (d).}
\label{Fig4}
\end{figure}

If $m<0$, we observe in the experiment stable square-wave oscillations of
period close to $\tau_{D}/n$. Figure \ref{Fig4}(b), (c), and (d) show
oscillations of period close to $\tau_{D}$, $\tau_{D}/2$, and $\tau_{D}/3$,
respectively. They are obtained by exciting the OEO\ with periodic signals of
different frequencies as described in the caption. Oscillations of period
close to $\tau_{D}/4$ are also observed. Moreover, we find stable
slowly-varying oscillations [See Fig. \ref{Fig4} (a)] in agreement with our
previous stability analysis that predicts a Hopf bifurcation for $m<0$ with a low
frequency. As for the case $m>0$, we do not find higher order harmonic
oscillations because of the bandwidth limitation of the pattern generator preventing us to initialize the system with frequencies above a certain threshold.%

\begin{figure}
\centering
\includegraphics[width=10cm]{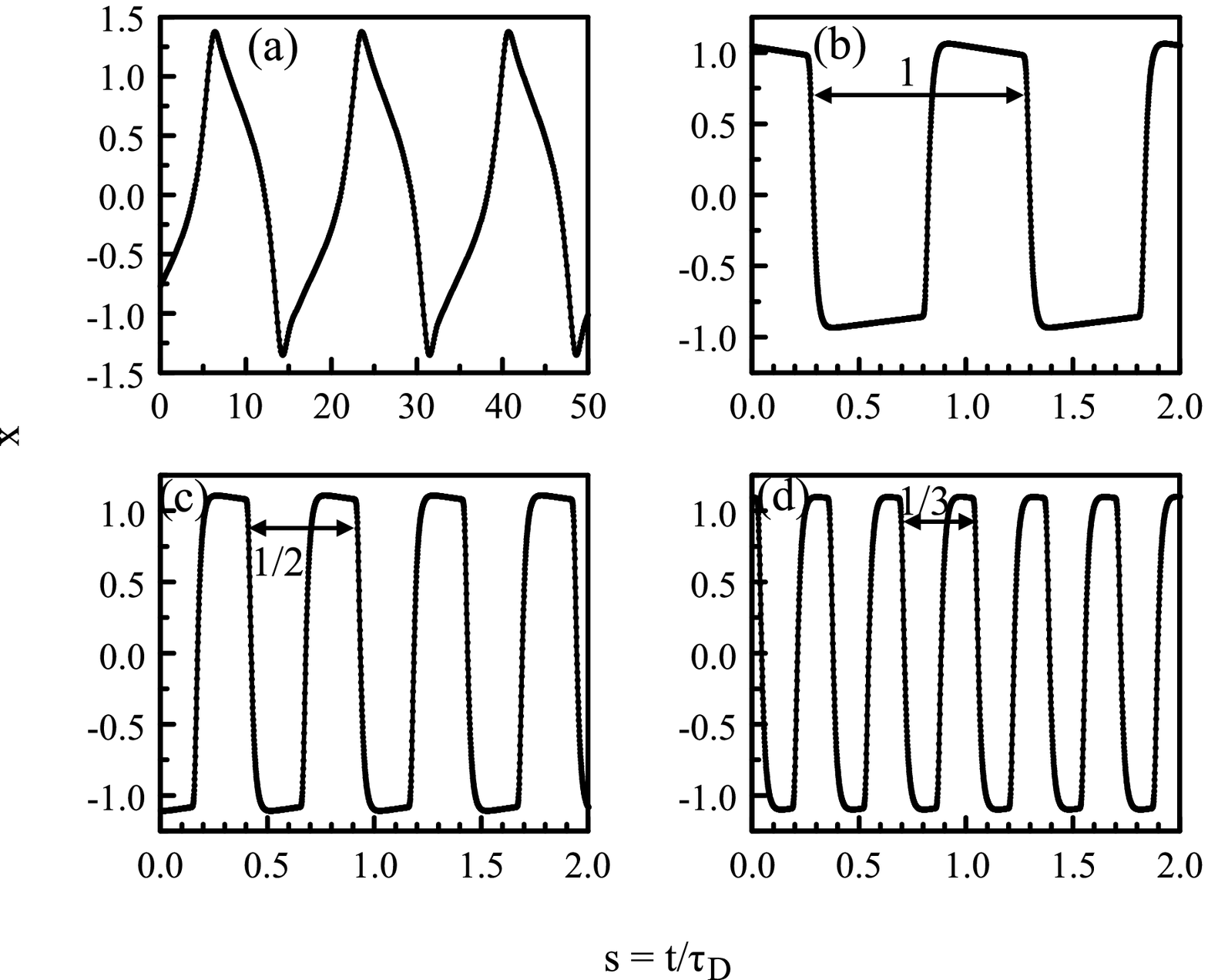}
\caption{Numerical time series obtained from Eqs. (\ref{Eq6}) and (\ref{Eq7}).
(a) Slowly varying solutions obtained with $x\left(  s\right)  =\cos\left(
0.5\pi s\right)  $ and $y\left(  s\right)  =0$ ($-1<s<0);$ (b) oscillations of
period close to $1$ with $x\left(  s\right)  =\cos\left(  2\pi s\right)  $ and
$y\left(  s\right)  =0$ ($-1<s<0)$; (c) oscillations of period close to $1/2$
obtained with $x\left(  s\right)  =\cos\left(  4\pi s\right)  $ and $y\left(
s\right)  =0$ ($-1<s<0);$ (d) oscillations of period close to $1/3$ with
$x\left(  s\right)  =\cos\left(  6\pi s\right)  $ and $y\left(  s\right)  =0$
($-1<s<0)$ The values of the control parameters are the same as in Fig.
\ref{Fig4}: $m=-0.845$ and $\beta=1.94$ for (a) and (b) and $m=-0.785$ and
$\beta=2.2$ for (c) and (d).}
\label{Fig5}%
\end{figure}

Integrating Eqs. (\ref{Eq6}) and (\ref{Eq7}) using different initial functions
leads to similar time-periodic regimes.\ The slowly-varying oscillations are
shown in Fig. \ref{Fig5}(a) and exhibit a period close to $T=17.2$.\ With the
Hopf bifurcation frequency $\omega_{0}$ given in (\ref{OE010}),\ we compute
$T_{0}=2\pi/\omega_{0}\simeq14$ which is of the same order of magnitude as
$T$. Figures \ref{Fig5}(b), (c), and (d) show oscillations of period close to
$1$, $1/2$, and $1/3$, respectively.

We investigate numerically the effect of changing $m<0$. The first Hopf bifurcation lines are shown in Fig. \ref{Duke100}(b). In contrast to the
case $m>0$ where the square-wave remains symmetric (same plateau
lengths), the shape of the square-wave depends here on $m$.\ If $m=-\pi/4$, we
observe symmetric square-wave oscillations with a period close to $\tau
_{D}/n.$ However, if $m+\pi/4\neq0,$ the square-wave becomes asymmetric with
different duty lengths for each plateau. The total period remains constant. As
for the case $m>0$, the square-wave oscillations become more robust if $\beta$
increases.\ The same properties are observed experimentally. Figure
\ref{Fig5}(b) and Figs.\ \ref{Fig5}(c) (d) are obtained using slightly
different values of the parameters $m$ and $\beta$. From Figs.\ \ref{Fig5}(c)
and (d), we find that the mean values of the plateaus are identical for the
two periodic regimes and are given by
\begin{equation}
x_{\max}\simeq1.1\text{ and }x_{\min}\simeq-1.1.\label{extrema1}%
\end{equation}

\section{Linear stability of the plateaus \label{stability}}%

\begin{figure}
\centering
\includegraphics[width=10cm]{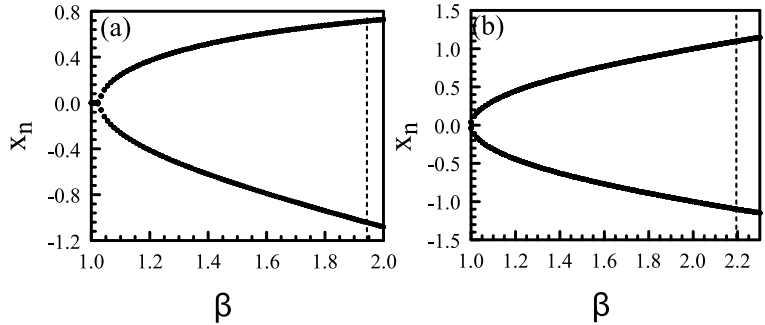}
\caption{Stable fixed points of Eq.\ (\ref{map1}).\ (a) $m=0.665$. The diagram
shows branches of a Period 2 fixed point. The dashed line corresponds to the
experimental and numerical value of $\beta$. (b) $m=-0.785$.\ The diagram
shows two branches of Period 1 fixed points. The dashed lines corresponds to
the experimental and numerical value of $\beta$.}
\label{Fig6}
\end{figure}
In the limit $\delta\rightarrow0$ and $\varepsilon\rightarrow0$, Eqs.
(\ref{Eq6}) and (\ref{Eq7}) reduce to a single equation for a map given by
\begin{equation}
x_{n+1}=\beta\left[  \cos^{2}\left(  m+\tanh(x_{n}\right)  )-\cos^{2}\left(
m\right)  \right]  .\label{map1}%
\end{equation}
Here, we demonstrate that the plateaus of the square-waves can be partially
understood by considering the stable fixed points of this map. For the case
$m>0$, there is a Hopf bifurcation at $\beta_{c}=1/\sin(2m)$ to a stable
period-$2$ fixed point [see Fig. \ref{Fig6}(a)]. For the values of $m$ and
$\beta$ used in our experiments and simulations, the diagram in Fig.
\ref{Fig6}(a) indicates
\begin{equation}
x_{\max}=0.72\text{ and }x_{\min}=-1.04,\label{map2}%
\end{equation}
which agree qualitatively with the values (\ref{extrema}) estimated from the numerical
simulations.\ 

For the case $m<0,$ Eq. (\ref{map1}) admits two stable period-1 fixed points
that appear at $\beta_{c}=-1/\sin(2m)$ [$x_{n}>0$ \ and $x_{n}<0,$ see Fig.
\ref{Fig6}(b)]. For the values of the parameters used in our experiments and
simulations [Figs.\ \ref{Fig4}(c)(d) and for Figs. \ref{Fig5}(c)(d)], the
diagram in Fig.\ \ref{Fig6}(b) indicates
\begin{equation}
x_{\max}=1.10\text{ and }x_{\min}-1.10,\label{map3}%
\end{equation}
which agree quantitatively with the values (\ref{extrema1}) obtained from the
numerical simulations.

\section{Discussion\label{conclusions}}

Symmetric and asymmetric square-wave oscillations have previously been found
\cite{PhysRevE.79.026208,weicker:12b,Rosin11,Weicker:13} and were related to
the first Hopf bifurcation of a basic steady state. Here, we concentrate on
the next primary Hopf bifurcations and show that they quickly stabilize above
critical amplitudes.\ This is the bifurcation scenario related to the Eckhaus
instability known to exist in spatially extended systems.\ The Eckhaus
instability has been predicted to occur in a simple model equation in the
limit of large delays \cite{PhysRevLett.96.220201}. The idea is based on the observation that all Hopf bifurcation points move to a critical value in the limit of large delay (in our case, $\varepsilon\rightarrow0$ and $\delta\rightarrow0$). We may then apply the method of multiple time scales and formulate a partial differential equation for a small amplitude solution. In \cite{PhysRevLett.96.220201}, a single variable complex Ginzburg-Landau equation was derived for which the stability of the different periodic solutions can be demonstrated analytically. The mechanism responsible for the stabilization of each branches of periodic solutions is called the Eckhaus instability. Assuming $\beta-1=O(\varepsilon^2)$ and $\delta=O(\varepsilon^2)$, we have found that two coupled partial differential equations can be derived from Eqs. (\ref{Eq6}) and (\ref{Eq7}). By contrast to the case studied in \cite{PhysRevLett.96.220201}, these equations cannot be solved analytically. However, their similitude to the Ginzburg-Landau equation suggests that distinct stable periodic solutions may coexist through the same Eckhaus scenario. In this paper, we demonstrated both experimentally and numerically that this coexistence of square-waves with distinct periods is possible. Those regimes have already been referenced
in previous studies of OEOs but not as coexisting solutions. They were
obtained as sequencial jumps when varying either the delay
\cite{Illing2005180,illing06} or the low frequency cutoff
\cite{Rosin11,Rosin_master_thesis}. Here, we showed how to obtain these regimes systematically in an experiment without varying any parameters of the
OEO system. A remarkable property of our OEO is the possibility to compare
quantitatively experimental observations and numerical simulations. It
motivates asymptotic studies of the OEO equations based on the large delay
limit \cite{Weicker:13}.

We believe that this multi-rhythmicity of square-waves resulting from nearby
Hopf bifurcations is generic to a large class of delay systems exhibiting a
large delay.\ Recently, we studied a semiconductor laser subject to polarization-rotated
feedback \cite{Friart:14} and found this coexistence of harmonic periodic
regimes both experimentally and numerically.\ The laser rate equations are
completely different from the dynamical equations for an OEO and are
mathematically more complex to analyze.\ The common property is the presence
of nearby Hopf bifurcation points leading to square-waves with periods that
are close to $2\tau_{D}/\left(  2n+1\right)  $, where $n=0,1,2,...$.

\begin{acknowledgments}
L.W. acknowledges the Belgian F.R.I.A., the Conseil Régional de Lorraine, the Agence Nationale de la Recherche (ANR) TINO project (ANR-12-JS03-005) and external funds from the F.N.R.S. T.E. acknowledges the support of the F.N.R.S. This work also benefited from the support of the Belgian Science Policy Office under Grant No IAP-7/35 “photonics@be.” D.P.R. and D.J.G. gratefully acknowledge the financial support by the U.S. Army Research Office, Grant No. W911NF-12-1-0099. 
\end{acknowledgments}

\bibliography{OEOs}

\begin{thebibliography}{29}%
\makeatletter
\providecommand \@ifxundefined [1]{%
 \@ifx{#1\undefined}
}%
\providecommand \@ifnum [1]{%
 \ifnum #1\expandafter \@firstoftwo
 \else \expandafter \@secondoftwo
 \fi
}%
\providecommand \@ifx [1]{%
 \ifx #1\expandafter \@firstoftwo
 \else \expandafter \@secondoftwo
 \fi
}%
\providecommand \natexlab [1]{#1}%
\providecommand \enquote  [1]{``#1''}%
\providecommand \bibnamefont  [1]{#1}%
\providecommand \bibfnamefont [1]{#1}%
\providecommand \citenamefont [1]{#1}%
\providecommand \href@noop [0]{\@secondoftwo}%
\providecommand \href [0]{\begingroup \@sanitize@url \@href}%
\providecommand \@href[1]{\@@startlink{#1}\@@href}%
\providecommand \@@href[1]{\endgroup#1\@@endlink}%
\providecommand \@sanitize@url [0]{\catcode `\\12\catcode `\$12\catcode
  `\&12\catcode `\#12\catcode `\^12\catcode `\_12\catcode `\%12\relax}%
\providecommand \@@startlink[1]{}%
\providecommand \@@endlink[0]{}%
\providecommand \url  [0]{\begingroup\@sanitize@url \@url }%
\providecommand \@url [1]{\endgroup\@href {#1}{\urlprefix }}%
\providecommand \urlprefix  [0]{URL }%
\providecommand \Eprint [0]{\href }%
\providecommand \doibase [0]{http://dx.doi.org/}%
\providecommand \selectlanguage [0]{\@gobble}%
\providecommand \bibinfo  [0]{\@secondoftwo}%
\providecommand \bibfield  [0]{\@secondoftwo}%
\providecommand \translation [1]{[#1]}%
\providecommand \BibitemOpen [0]{}%
\providecommand \bibitemStop [0]{}%
\providecommand \bibitemNoStop [0]{.\EOS\space}%
\providecommand \EOS [0]{\spacefactor3000\relax}%
\providecommand \BibitemShut  [1]{\csname bibitem#1\endcsname}%
\let\auto@bib@innerbib\@empty
%</preamble>
\bibitem [{\citenamefont {Erneux}\ and\ \citenamefont
  {Glorieux}(2010)}]{erneux2010laser}%
  \BibitemOpen
  \bibfield  {author} {\bibinfo {author} {\bibfnamefont {T.}~\bibnamefont
  {Erneux}}\ and\ \bibinfo {author} {\bibfnamefont {P.}~\bibnamefont
  {Glorieux}},\ }\href@noop {} {\emph {\bibinfo {title} {Laser dynamics}}}\
  (\bibinfo  {publisher} {Cambridge University Press},\ \bibinfo {address}
  {Cambridge, UK},\ \bibinfo {year} {2010})\BibitemShut {NoStop}%
\bibitem [{\citenamefont {Erneux}(2009)}]{erneux}%
  \BibitemOpen
  \bibfield  {author} {\bibinfo {author} {\bibfnamefont {T.}~\bibnamefont
  {Erneux}},\ }\href@noop {} {\emph {\bibinfo {title} {Applied Delay
  Differential Equations}}}\ (\bibinfo  {publisher} {Springer},\ \bibinfo
  {address} {New York},\ \bibinfo {year} {2009})\BibitemShut {NoStop}%
\bibitem [{\citenamefont {Balachandran}\ \emph {et~al.}(2009)\citenamefont
  {Balachandran}, \citenamefont {Kam\'{a}r-Nagy}, \citenamefont {Gilsinn},\
  and\ \citenamefont {David}}]{bala}%
  \BibitemOpen
  \bibfield  {author} {\bibinfo {author} {\bibfnamefont {A.}~\bibnamefont
  {Balachandran}}, \bibinfo {author} {\bibfnamefont {T.}~\bibnamefont
  {Kam\'{a}r-Nagy}}, \bibinfo {author} {\bibfnamefont {D.}~\bibnamefont
  {Gilsinn}}, \ and\ \bibinfo {author} {\bibfnamefont {E.}~\bibnamefont
  {David}},\ }\href@noop {} {\emph {\bibinfo {title} {Delay Differential
  Equations, Recent Advances and New Directions}}}\ (\bibinfo  {publisher}
  {Springer},\ \bibinfo {address} {New York},\ \bibinfo {year}
  {2009})\BibitemShut {NoStop}%
\bibitem [{\citenamefont {Stepan}(2009)}]{Stepan28032009}%
  \BibitemOpen
  \bibfield  {author} {\bibinfo {author} {\bibfnamefont {G.}~\bibnamefont
  {Stepan}},\ }\href
  {http://rsta.royalsocietypublishing.org/content/367/1891/1059.short}
  {\bibfield  {journal} {\bibinfo  {journal} {Philosophical Transactions of the
  Royal Society A: Mathematical, Physical and Engineering Sciences}\ }\textbf
  {\bibinfo {volume} {367}},\ \bibinfo {pages} {1059} (\bibinfo {year}
  {2009})}\BibitemShut {NoStop}%
\bibitem [{\citenamefont {Atay}(2010)}]{atay2010complex}%
  \BibitemOpen
  \bibfield  {author} {\bibinfo {author} {\bibfnamefont {F.}~\bibnamefont
  {Atay}},\ }\href@noop {} {\emph {\bibinfo {title} {Complex time-delay
  systems}}}\ (\bibinfo  {publisher} {Springer},\ \bibinfo {address} {New
  York},\ \bibinfo {year} {2010})\BibitemShut {NoStop}%
\bibitem [{\citenamefont {Just}\ \emph {et~al.}(2010)\citenamefont {Just},
  \citenamefont {Pelster}, \citenamefont {Schanz},\ and\ \citenamefont
  {Sch{\"o}ll}}]{just2010delayed}%
  \BibitemOpen
  \bibfield  {author} {\bibinfo {author} {\bibfnamefont {W.}~\bibnamefont
  {Just}}, \bibinfo {author} {\bibfnamefont {A.}~\bibnamefont {Pelster}},
  \bibinfo {author} {\bibfnamefont {M.}~\bibnamefont {Schanz}}, \ and\ \bibinfo
  {author} {\bibfnamefont {E.}~\bibnamefont {Sch{\"o}ll}},\ }\href@noop {}
  {\bibfield  {journal} {\bibinfo  {journal} {Philosophical Transactions of the
  Royal Society A: Mathematical, Physical and Engineering Sciences}\ }\textbf
  {\bibinfo {volume} {368}},\ \bibinfo {pages} {303} (\bibinfo {year}
  {2010})}\BibitemShut {NoStop}%
\bibitem [{\citenamefont {Kalm\'{a}r-Nagy}\ \emph {et~al.}(2010)\citenamefont
  {Kalm\'{a}r-Nagy}, \citenamefont {Olgac},\ and\ \citenamefont
  {St\'{e}p\'{a}n}}]{KalmarNagy01062010}%
  \BibitemOpen
  \bibfield  {author} {\bibinfo {author} {\bibfnamefont {T.}~\bibnamefont
  {Kalm\'{a}r-Nagy}}, \bibinfo {author} {\bibfnamefont {N.}~\bibnamefont
  {Olgac}}, \ and\ \bibinfo {author} {\bibfnamefont {G.}~\bibnamefont
  {St\'{e}p\'{a}n}},\ }\href {\doibase 10.1177/1077546309341138} {\bibfield
  {journal} {\bibinfo  {journal} {Journal of Vibration and Control}\ }\textbf
  {\bibinfo {volume} {16}},\ \bibinfo {pages} {941} (\bibinfo {year} {2010})},\
  \Eprint {http://arxiv.org/abs/http://jvc.sagepub.com/cgi/reprint/16/7-8/941}
  {http://jvc.sagepub.com/cgi/reprint/16/7-8/941} \BibitemShut {NoStop}%
\bibitem [{\citenamefont {Lakshmanan}\ and\ \citenamefont
  {Senthilkumar}(2011)}]{lakshmanan2011dynamics}%
  \BibitemOpen
  \bibfield  {author} {\bibinfo {author} {\bibfnamefont {M.}~\bibnamefont
  {Lakshmanan}}\ and\ \bibinfo {author} {\bibfnamefont {D.}~\bibnamefont
  {Senthilkumar}},\ }\href@noop {} {\emph {\bibinfo {title} {Dynamics of
  nonlinear time-delay systems}}}\ (\bibinfo  {publisher} {Springer},\ \bibinfo
  {address} {New York},\ \bibinfo {year} {2011})\BibitemShut {NoStop}%
\bibitem [{\citenamefont {Smith}(2011)}]{smith2011introduction}%
  \BibitemOpen
  \bibfield  {author} {\bibinfo {author} {\bibfnamefont {H.}~\bibnamefont
  {Smith}},\ }\href@noop {} {\emph {\bibinfo {title} {An introduction to delay
  differential equations with applications to the life sciences}}}\ (\bibinfo
  {publisher} {Springer},\ \bibinfo {address} {New York},\ \bibinfo {year}
  {2011})\BibitemShut {NoStop}%
\bibitem [{\citenamefont {Soriano}\ \emph {et~al.}(2013)\citenamefont
  {Soriano}, \citenamefont {Garc{\'\i}a-Ojalvo}, \citenamefont {Mirasso},\ and\
  \citenamefont {Fischer}}]{soriano2013complex}%
  \BibitemOpen
  \bibfield  {author} {\bibinfo {author} {\bibfnamefont {M.~C.}\ \bibnamefont
  {Soriano}}, \bibinfo {author} {\bibfnamefont {J.}~\bibnamefont
  {Garc{\'\i}a-Ojalvo}}, \bibinfo {author} {\bibfnamefont {C.~R.}\ \bibnamefont
  {Mirasso}}, \ and\ \bibinfo {author} {\bibfnamefont {I.}~\bibnamefont
  {Fischer}},\ }\href
  {http://journals.aps.org/rmp/abstract/10.1103/RevModPhys.85.421} {\bibfield
  {journal} {\bibinfo  {journal} {Reviews of Modern Physics}\ }\textbf
  {\bibinfo {volume} {85}},\ \bibinfo {pages} {421} (\bibinfo {year}
  {2013})}\BibitemShut {NoStop}%
\bibitem [{\citenamefont {Larger}(2013)}]{Larger28092013}%
  \BibitemOpen
  \bibfield  {author} {\bibinfo {author} {\bibfnamefont {L.}~\bibnamefont
  {Larger}},\ }\href
  {http://rsta.royalsocietypublishing.org/content/371/1999/20120464.short}
  {\bibfield  {journal} {\bibinfo  {journal} {Philosophical Transactions of the
  Royal Society A: Mathematical, Physical and Engineering Sciences}\ }\textbf
  {\bibinfo {volume} {371}},\ \bibinfo {pages} {20120464} (\bibinfo {year}
  {2013})}\BibitemShut {NoStop}%
\bibitem [{\citenamefont {Devgan}(2013)}]{devgan2013review}%
  \BibitemOpen
  \bibfield  {author} {\bibinfo {author} {\bibfnamefont {P.}~\bibnamefont
  {Devgan}},\ }\href@noop {} {\bibfield  {journal} {\bibinfo  {journal}
  {International Scholarly Research Notices}\ }\textbf {\bibinfo {volume}
  {2013}} (\bibinfo {year} {2013})}\BibitemShut {NoStop}%
\bibitem [{\citenamefont {Chembo}\ \emph {et~al.}(2009)\citenamefont {Chembo},
  \citenamefont {Hmima}, \citenamefont {Lacourt}, \citenamefont {Larger},\ and\
  \citenamefont {Dudley}}]{yanne:09}%
  \BibitemOpen
  \bibfield  {author} {\bibinfo {author} {\bibfnamefont {Y.~K.}\ \bibnamefont
  {Chembo}}, \bibinfo {author} {\bibfnamefont {A.}~\bibnamefont {Hmima}},
  \bibinfo {author} {\bibfnamefont {P.-A.}\ \bibnamefont {Lacourt}}, \bibinfo
  {author} {\bibfnamefont {L.}~\bibnamefont {Larger}}, \ and\ \bibinfo {author}
  {\bibfnamefont {J.~M.}\ \bibnamefont {Dudley}},\ }\href
  {http://ieeexplore.ieee.org/xpls/abs_all.jsp?arnumber=5169948} {\bibfield
  {journal} {\bibinfo  {journal} {Lightwave Technology, Journal of}\ }\textbf
  {\bibinfo {volume} {27}},\ \bibinfo {pages} {5160} (\bibinfo {year}
  {2009})}\BibitemShut {NoStop}%
\bibitem [{\citenamefont {Gastaud}\ \emph {et~al.}(2004)\citenamefont
  {Gastaud}, \citenamefont {Poinsot}, \citenamefont {Larger}, \citenamefont
  {Merolla}, \citenamefont {Hanna}, \citenamefont {Goedgebuer},\ and\
  \citenamefont {Malassenet}}]{gastaud04}%
  \BibitemOpen
  \bibfield  {author} {\bibinfo {author} {\bibfnamefont {N.}~\bibnamefont
  {Gastaud}}, \bibinfo {author} {\bibfnamefont {S.}~\bibnamefont {Poinsot}},
  \bibinfo {author} {\bibfnamefont {L.}~\bibnamefont {Larger}}, \bibinfo
  {author} {\bibfnamefont {J.}~\bibnamefont {Merolla}}, \bibinfo {author}
  {\bibfnamefont {M.}~\bibnamefont {Hanna}}, \bibinfo {author} {\bibfnamefont
  {J.}~\bibnamefont {Goedgebuer}}, \ and\ \bibinfo {author} {\bibfnamefont
  {F.}~\bibnamefont {Malassenet}},\ }\href
  {http://digital-library.theiet.org/content/journals/10.1049/el_20045072}
  {\bibfield  {journal} {\bibinfo  {journal} {Electronics letters}\ }\textbf
  {\bibinfo {volume} {40}},\ \bibinfo {pages} {898} (\bibinfo {year}
  {2004})}\BibitemShut {NoStop}%
\bibitem [{\citenamefont {Callan}\ \emph {et~al.}(2010)\citenamefont {Callan},
  \citenamefont {Illing}, \citenamefont {Gao}, \citenamefont {Gauthier},\ and\
  \citenamefont {Sch{\"o}ll}}]{callan}%
  \BibitemOpen
  \bibfield  {author} {\bibinfo {author} {\bibfnamefont {K.~E.}\ \bibnamefont
  {Callan}}, \bibinfo {author} {\bibfnamefont {L.}~\bibnamefont {Illing}},
  \bibinfo {author} {\bibfnamefont {Z.}~\bibnamefont {Gao}}, \bibinfo {author}
  {\bibfnamefont {D.~J.}\ \bibnamefont {Gauthier}}, \ and\ \bibinfo {author}
  {\bibfnamefont {E.}~\bibnamefont {Sch{\"o}ll}},\ }\href
  {http://journals.aps.org/prl/abstract/10.1103/PhysRevLett.104.113901}
  {\bibfield  {journal} {\bibinfo  {journal} {Physical review letters}\
  }\textbf {\bibinfo {volume} {104}},\ \bibinfo {pages} {113901} (\bibinfo
  {year} {2010})}\BibitemShut {NoStop}%
\bibitem [{\citenamefont {Peil}\ \emph {et~al.}(2009)\citenamefont {Peil},
  \citenamefont {Jacquot}, \citenamefont {Chembo}, \citenamefont {Larger},\
  and\ \citenamefont {Erneux}}]{PhysRevE.79.026208}%
  \BibitemOpen
  \bibfield  {author} {\bibinfo {author} {\bibfnamefont {M.}~\bibnamefont
  {Peil}}, \bibinfo {author} {\bibfnamefont {M.}~\bibnamefont {Jacquot}},
  \bibinfo {author} {\bibfnamefont {Y.~K.}\ \bibnamefont {Chembo}}, \bibinfo
  {author} {\bibfnamefont {L.}~\bibnamefont {Larger}}, \ and\ \bibinfo {author}
  {\bibfnamefont {T.}~\bibnamefont {Erneux}},\ }\href
  {http://journals.aps.org/pre/abstract/10.1103/PhysRevE.79.026208} {\bibfield
  {journal} {\bibinfo  {journal} {Physical Review E}\ }\textbf {\bibinfo
  {volume} {79}},\ \bibinfo {pages} {026208} (\bibinfo {year}
  {2009})}\BibitemShut {NoStop}%
\bibitem [{\citenamefont {Levy}\ \emph {et~al.}(2009)\citenamefont {Levy},
  \citenamefont {Horowitz},\ and\ \citenamefont {Menyuk}}]{Levy:09}%
  \BibitemOpen
  \bibfield  {author} {\bibinfo {author} {\bibfnamefont {E.~C.}\ \bibnamefont
  {Levy}}, \bibinfo {author} {\bibfnamefont {M.}~\bibnamefont {Horowitz}}, \
  and\ \bibinfo {author} {\bibfnamefont {C.~R.}\ \bibnamefont {Menyuk}},\
  }\href {http://www.opticsinfobase.org/abstract.cfm?uri=josab-26-1-148}
  {\bibfield  {journal} {\bibinfo  {journal} {JOSA B}\ }\textbf {\bibinfo
  {volume} {26}},\ \bibinfo {pages} {148} (\bibinfo {year} {2009})}\BibitemShut
  {NoStop}%
\bibitem [{\citenamefont {Weicker}\ \emph {et~al.}(2012)\citenamefont
  {Weicker}, \citenamefont {Erneux}, \citenamefont {d'Huys}, \citenamefont
  {Danckaert}, \citenamefont {Jacquot}, \citenamefont {Chembo},\ and\
  \citenamefont {Larger}}]{weicker:12b}%
  \BibitemOpen
  \bibfield  {author} {\bibinfo {author} {\bibfnamefont {L.}~\bibnamefont
  {Weicker}}, \bibinfo {author} {\bibfnamefont {T.}~\bibnamefont {Erneux}},
  \bibinfo {author} {\bibfnamefont {O.}~\bibnamefont {d'Huys}}, \bibinfo
  {author} {\bibfnamefont {J.}~\bibnamefont {Danckaert}}, \bibinfo {author}
  {\bibfnamefont {M.}~\bibnamefont {Jacquot}}, \bibinfo {author} {\bibfnamefont
  {Y.}~\bibnamefont {Chembo}}, \ and\ \bibinfo {author} {\bibfnamefont
  {L.}~\bibnamefont {Larger}},\ }\href
  {http://journals.aps.org/pre/abstract/10.1103/PhysRevE.86.055201} {\bibfield
  {journal} {\bibinfo  {journal} {Physical Review E}\ }\textbf {\bibinfo
  {volume} {86}},\ \bibinfo {pages} {055201} (\bibinfo {year}
  {2012})}\BibitemShut {NoStop}%
\bibitem [{\citenamefont {Wolfrum}\ and\ \citenamefont
  {Yanchuk}(2006)}]{PhysRevLett.96.220201}%
  \BibitemOpen
  \bibfield  {author} {\bibinfo {author} {\bibfnamefont {M.}~\bibnamefont
  {Wolfrum}}\ and\ \bibinfo {author} {\bibfnamefont {S.}~\bibnamefont
  {Yanchuk}},\ }\href
  {http://journals.aps.org/prl/abstract/10.1103/PhysRevLett.96.220201}
  {\bibfield  {journal} {\bibinfo  {journal} {Physical review letters}\
  }\textbf {\bibinfo {volume} {96}},\ \bibinfo {pages} {220201} (\bibinfo
  {year} {2006})}\BibitemShut {NoStop}%
\bibitem [{\citenamefont {Tuckerman}\ and\ \citenamefont
  {Barkley}(1990)}]{Tuckerman199057}%
  \BibitemOpen
  \bibfield  {author} {\bibinfo {author} {\bibfnamefont {L.~S.}\ \bibnamefont
  {Tuckerman}}\ and\ \bibinfo {author} {\bibfnamefont {D.}~\bibnamefont
  {Barkley}},\ }\href
  {http://www.sciencedirect.com/science/article/pii/0167278990901134}
  {\bibfield  {journal} {\bibinfo  {journal} {Physica D: Nonlinear Phenomena}\
  }\textbf {\bibinfo {volume} {46}},\ \bibinfo {pages} {57} (\bibinfo {year}
  {1990})}\BibitemShut {NoStop}%
\bibitem [{\citenamefont {Illing}\ and\ \citenamefont
  {Gauthier}(2005)}]{Illing2005180}%
  \BibitemOpen
  \bibfield  {author} {\bibinfo {author} {\bibfnamefont {L.}~\bibnamefont
  {Illing}}\ and\ \bibinfo {author} {\bibfnamefont {D.~J.}\ \bibnamefont
  {Gauthier}},\ }\href
  {http://www.sciencedirect.com/science/article/pii/S0167278905003118}
  {\bibfield  {journal} {\bibinfo  {journal} {Physica D: Nonlinear Phenomena}\
  }\textbf {\bibinfo {volume} {210}},\ \bibinfo {pages} {180} (\bibinfo {year}
  {2005})}\BibitemShut {NoStop}%
\bibitem [{\citenamefont {Illing}\ and\ \citenamefont
  {Gauthier}(2006)}]{illing06}%
  \BibitemOpen
  \bibfield  {author} {\bibinfo {author} {\bibfnamefont {L.}~\bibnamefont
  {Illing}}\ and\ \bibinfo {author} {\bibfnamefont {D.~J.}\ \bibnamefont
  {Gauthier}},\ }\href
  {http://scitation.aip.org/content/aip/journal/chaos/16/3/10.1063/1.2335814}
  {\bibfield  {journal} {\bibinfo  {journal} {Chaos: An Interdisciplinary
  Journal of Nonlinear Science}\ }\textbf {\bibinfo {volume} {16}},\ \bibinfo
  {pages} {033119} (\bibinfo {year} {2006})}\BibitemShut {NoStop}%
\bibitem [{\citenamefont {Rosin}\ \emph {et~al.}(2011)\citenamefont {Rosin},
  \citenamefont {Callan}, \citenamefont {Gauthier},\ and\ \citenamefont
  {Sch{\"o}ll}}]{Rosin11}%
  \BibitemOpen
  \bibfield  {author} {\bibinfo {author} {\bibfnamefont {D.~P.}\ \bibnamefont
  {Rosin}}, \bibinfo {author} {\bibfnamefont {K.~E.}\ \bibnamefont {Callan}},
  \bibinfo {author} {\bibfnamefont {D.~J.}\ \bibnamefont {Gauthier}}, \ and\
  \bibinfo {author} {\bibfnamefont {E.}~\bibnamefont {Sch{\"o}ll}},\ }\href
  {http://iopscience.iop.org/0295-5075/96/3/34001} {\bibfield  {journal}
  {\bibinfo  {journal} {EPL (Europhysics Letters)}\ }\textbf {\bibinfo {volume}
  {96}},\ \bibinfo {pages} {34001} (\bibinfo {year} {2011})}\BibitemShut
  {NoStop}%
\bibitem [{\citenamefont {Rosin}(2011)}]{Rosin_master_thesis}%
  \BibitemOpen
  \bibfield  {author} {\bibinfo {author} {\bibfnamefont {D.}~\bibnamefont
  {Rosin}},\ }\emph {\bibinfo {title} {Pulse train solutions in a time-delayed
  opto-electronic oscillator}},\ \href@noop {} {Master's thesis},\ \bibinfo
  {school} {Technische Universit\"at Berlin} (\bibinfo {year}
  {2011})\BibitemShut {NoStop}%
\bibitem [{\citenamefont {Kouomou}\ \emph {et~al.}(2005)\citenamefont
  {Kouomou}, \citenamefont {Colet}, \citenamefont {Larger},\ and\ \citenamefont
  {Gastaud}}]{PhysRevLett.95.203903}%
  \BibitemOpen
  \bibfield  {author} {\bibinfo {author} {\bibfnamefont {Y.~C.}\ \bibnamefont
  {Kouomou}}, \bibinfo {author} {\bibfnamefont {P.}~\bibnamefont {Colet}},
  \bibinfo {author} {\bibfnamefont {L.}~\bibnamefont {Larger}}, \ and\ \bibinfo
  {author} {\bibfnamefont {N.}~\bibnamefont {Gastaud}},\ }\href
  {http://journals.aps.org/prl/abstract/10.1103/PhysRevLett.95.203903}
  {\bibfield  {journal} {\bibinfo  {journal} {Physical review letters}\
  }\textbf {\bibinfo {volume} {95}},\ \bibinfo {pages} {203903} (\bibinfo
  {year} {2005})}\BibitemShut {NoStop}%
\bibitem [{Note1()}]{Note1}%
  \BibitemOpen
  \bibinfo {note} {From Eqs. (1) and (2) in \cite {callan}, we obtain our Eqs.
  (1) and (2) with $\varepsilon = (T\Delta )^{-1} = (T(\omega _{+} - \omega
  _{-}))^{-1}$, and $\delta =T\omega _{0}^{2}/\Delta =T(\omega _{+}\omega
  _{-})/(\omega _{+}-\omega _{-})$ and $\beta =\gamma $. $T$ is the delay of
  the feedback loop, $\omega _{-}$ and $\omega _{+}$ represent the low and high
  frequency cut-off of the bandpass filter, respectively. Since $\omega _{+}
  \gg \omega _{-}$, $\varepsilon \simeq (T \omega _{+})^{-1}$ and $\delta
  \simeq T\omega _{-}$}\BibitemShut {NoStop}%
\bibitem [{Note2()}]{Note2}%
  \BibitemOpen
  \bibinfo {note} {From Eqs. (4) and (5) in \cite {PhysRevLett.95.203903},
  introduce $\omega \DOTSB \protect \relbar \protect \joinrel \rightarrow
  \omega R,$ $\delta \DOTSB \protect \relbar \protect \joinrel \rightarrow
  \varepsilon R,$ $\varepsilon \DOTSB \protect \relbar \protect \joinrel
  \rightarrow 1/R.$ The new parameters $\varepsilon $ and $\delta $ are small
  and their values are given in Section \ref {intro}.}\BibitemShut {Stop}%
\bibitem [{\citenamefont {Weicker}\ \emph {et~al.}(2013)\citenamefont
  {Weicker}, \citenamefont {Erneux}, \citenamefont {D'Huys}, \citenamefont
  {Danckaert}, \citenamefont {Jacquot}, \citenamefont {Chembo},\ and\
  \citenamefont {Larger}}]{Weicker:13}%
  \BibitemOpen
  \bibfield  {author} {\bibinfo {author} {\bibfnamefont {L.}~\bibnamefont
  {Weicker}}, \bibinfo {author} {\bibfnamefont {T.}~\bibnamefont {Erneux}},
  \bibinfo {author} {\bibfnamefont {O.}~\bibnamefont {D'Huys}}, \bibinfo
  {author} {\bibfnamefont {J.}~\bibnamefont {Danckaert}}, \bibinfo {author}
  {\bibfnamefont {M.}~\bibnamefont {Jacquot}}, \bibinfo {author} {\bibfnamefont
  {Y.}~\bibnamefont {Chembo}}, \ and\ \bibinfo {author} {\bibfnamefont
  {L.}~\bibnamefont {Larger}},\ }\href
  {http://rsta.royalsocietypublishing.org/content/371/1999/20120459.short}
  {\bibfield  {journal} {\bibinfo  {journal} {Philosophical Transactions of the
  Royal Society A: Mathematical, Physical and Engineering Sciences}\ }\textbf
  {\bibinfo {volume} {371}},\ \bibinfo {pages} {20120459} (\bibinfo {year}
  {2013})}\BibitemShut {NoStop}%
\bibitem [{\citenamefont {Friart}\ \emph {et~al.}(2014)\citenamefont {Friart},
  \citenamefont {Weicker}, \citenamefont {Danckaert},\ and\ \citenamefont
  {Erneux}}]{Friart:14}%
  \BibitemOpen
  \bibfield  {author} {\bibinfo {author} {\bibfnamefont {G.}~\bibnamefont
  {Friart}}, \bibinfo {author} {\bibfnamefont {L.}~\bibnamefont {Weicker}},
  \bibinfo {author} {\bibfnamefont {J.}~\bibnamefont {Danckaert}}, \ and\
  \bibinfo {author} {\bibfnamefont {T.}~\bibnamefont {Erneux}},\ }\href@noop {}
  {\bibfield  {journal} {\bibinfo  {journal} {Optics express}\ }\textbf
  {\bibinfo {volume} {22}},\ \bibinfo {pages} {6905} (\bibinfo {year}
  {2014})}\BibitemShut {NoStop}%
\end{thebibliography}%

\end{document}